\documentclass[runningheads]{llncs}
\usepackage[ruled,lined,linesnumbered]{algorithm2e} 

\usepackage{amsmath,amssymb,amsfonts}
\usepackage{graphicx}
\usepackage{booktabs} 
\usepackage[misc]{ifsym}
\usepackage{amsmath}
\usepackage{amsfonts}
\usepackage{multirow}
\usepackage{hhline}

\def\cR{{\mathcal R}}
\def \CTVM{{\mathsf{CTVM}}}
\def \IBS{{\mathsf{IBS}}}
\def \IBA{{\mathsf{IBA}}}
\def \BCT{{\mathsf{BCT}}}
\def \IVM{{\mathsf{IVM}}}
\def \TIPTOP{{\mathsf{TIPTOP}}}
\def \SSA{{\mathsf{SSA}}}
\def \DSSA{{\mathsf{DSSA}}}
\def \IC{{\mathsf{IC}}}
\def \LT{{\mathsf{LT}}}
\def \Cov{{\mathsf{Cov}}}
\def \Var{{\mathsf{Var}}}
\def \OPT{{\mathsf{OPT}}}

\def \IM{\mathsf{IM}}

\def \IGA{{\mathsf{IGA}}}

\def\hB{{\hat{\mathbb B}}}

\def\eE{{\mathbb E}}

\def\eI{{\mathbb I}}
\def\eB{{\mathbb B}}

\begin{document}

	\title{Cost-aware Targeted Viral Marketing: Approximation with Less Samples}
	\titlerunning{Cost-aware Targeted Viral Marketing: Approximation with Less Samples}
	%
	\author{Canh V. Pham \inst{1,2}\and
		Hieu V. Duong \inst{1}\and
		My T. Thai \inst{3}}
	\authorrunning{Canh V. Pham et al.}
	%
	\institute{People's Security Academy, Hanoi, Vietnam 
		\\
		\email{maicanhki@gmail.com, dvhieubg95@gmail.com}
		\and
		University of Engineering and Technology
		\\
		Vietnam National University, Hanoi,  Vietnam
		\and
		Dept. of Comp. \& Info. Sci. \& Eng., University of Florida, Gainesville, USA
		\\
		\email{mythai@cise.ufl.edu}
	}
	\maketitle              
	\begin{abstract}
		Cost-aware Targeted Viral Marketing ($\CTVM$), a generalization of Influence Maximization ($\IM$), has received a lot of attentions recently due to its commercial values. Previous approximation algorithms for this problem required a large number of samples to ensure approximate guarantee. In this paper, we propose an efficient approximation algorithm which uses fewer samples but provides the same theoretical guarantees based on generating and using important samples in its operation. Experiments on real social networks show that our proposed method outperforms the state-of-the-art algorithm which provides the same approximation ratio in terms of the number of required samples and running time.
		
	\end{abstract}
	\newtheorem{coro}{Corollary}
	\section{Introduction}
	Many companies have leveraged the ``word-of-mouth'' effect of Online social networks (OSNs) to conduct their marketing and advertising activities.
	One of the key problems in viral marketing is Influence Maximization ($\IM$), which aims at selecting a set of $k$ users, called a seed set, in a social network so that the expected number of influenced nodes is maximized. 
	Kempe {\em et al.} \cite{kem03} introduced the $\IM$ problem under two classical information diffusion models, namely, Independent Cascade ($\IC$) and Linear Threshold ($\LT$). The authors proved that $\IM$ is an NP-hard problem and designed a $(1-1/e)$-approximation algorithm based on the monotonic and submodular property of the influence function. Due to its immense application potential, a vast amount of work focused on $\IM$ in many respects: designing effective algorithms \cite{borg,nguyen_sigmod16,tang14,tang15} and studying variants with marketing applications \cite{cikm_12,least_cost_ToN,cost_effective,pit_joco14,bcim_as,zhang2014recent,wu2016mining}. 
	
	Borgs {\em et al.} \cite{borg} make a theoretical breakthrough by proposing a reverse influence sketch (RIS) algorithm which is the foundation for later efficient algorithms. This algorithm captures the influences in a reverse manner and guarantee $(1-1/e)$-approximation solution with probability at least $1-n^{-1}$ where $n$ is the number of nodes in the network. In a sequential work, Tang {\em et al.}  \cite{tang14}  first presented a $(1-1/e)$-approximation  algorithm that is scalable for billion-size networks with the running time reduces to $O((k + l)(m + n) \ln n \epsilon^{-2})$. \cite{tang15} later proposed the IMM algorithm, which further reduced the number of samples of RIS process by using Martingale analysis. Nguyen {\em et al.} \cite{nguyen_sigmod16,dssa_revise} 
	$\SSA/\DSSA$ algorithms to further reduce the running time up to
	orders of magnitude by modifying the original RIS framework.
	The algorithm keeps generating samples
	and stops at exponential check points to verify (stare) the quality solution. Recently, \cite{importance_sketch} indicates that it is possible to further reduce the number of samples of the above algorithms based on the use of important samples.
	
	In a more realistic scenario with taking into account both arbitrary
	cost for selecting a node and arbitrary benefit for influencing
	a node, Nguyen {\em et al.} \cite{ctvm_infocom} studied Cost-aware Targeted Viral Marketing ($\CTVM$) problem, a generalization of $\IM$, which aims to select a seed set within limited budget so that the expected total benefit over the influenced nodes (benefit function) is maximized. In this work, the benefit function can be estimated through benefit samples sampling algorithm, a generalized version of RIS and the authors proposed $\BCT$ algorithm, a $(1-1/\sqrt{e} -\epsilon)$-approximation algorithm with high probability with the number of required samples at least $O(n\ln(n^{k_{max}}/\delta)\epsilon^{-2})$ under $\IC$ model. 
	In another direction, Li {\em et al.} \cite{tiptop_infocom} solved $\CTVM$ with an almost exact algorithm $\TIPTOP$ approach, which can return the solution within a ratio of $(1-\epsilon)$ with high probability. The algorithm needs at most $O(\frac{nk\log n}{\OPT_k \epsilon^2})$ samples and no bound on the time complexity as this is an exact approach, not the approximation algorithm approach. However, the authors have shown that $\TIPTOP$ can run on the Twitter datasets within four hours \cite{tiptop_infocom}.
	
	In this paper, we tackle $\CTVM$ via an approximation approach (not exact) with a goal of obtaining the same approximation ratio $(1-1/\sqrt{e}-\epsilon)$ as in \cite{ctvm_infocom}, but significantly reducing the number of samples to $O( \frac{\rho n\log(M/\delta)}{\epsilon^2 \OPT})$ samples for $\rho<1$. Our algorithm, namely Importance sample-based for Viral Marketing ($\IVM$), contains two innovative techniques: 1) We note that importance samples (in the space of all benefit samples) can be used to estimate the benefit function. This leads to a general result of using importance sketches to estimate the influence spread function for $\IM$ \cite{importance_sketch}. 2) Base on that we design a new strategy to check approximation guarantee condition of candidate solutions. We develop two lower and upper bound functions to check approximation guarantee condition  and adequate statistical evidence on the solution quality for termination. Our algorithm takes lower total samples than $\BCT$, which is state of the art method with same approximation guarantee in both theoretical analysis and practical. 
	In summary, the contributions of this paper are as follows.
	\begin{itemize}
		\item We first present Importance Benefit Sample ($\IBS$) and 
		Importance Benefit Sampling Algorithm ($\IBA$), an algorithm to generate $\IBS$. We then show that the benefit function can be estimated through $\IBS$ (Lemma \ref{lem: ibs_est}). 
		\item We proposed $\IVM$, an efficient approximation algorithm which returns $(1-1/\sqrt{e}-\epsilon)$ with high probability and the expected number of required samples generated by our algorithm is $O( \frac{\rho n\log(M/\delta)}{\epsilon^2 \OPT})$, where $\rho<1$, and $M=k_{max}\binom{n}{k_0}$ under $\IC$ model. This is a better result than $\BCT$ algorithm.
		\item We conduct experiments on various real social
		networks. The experiments on some social networks suggest that $\IVM$ better than $\BCT$, a  current best method on $\CTVM$ with the same approximation guarantee, in terms of running time, number of required samples, and used memory. It achieves up to 153 times speed-up and the total required  samples less than 112 times than that of $\BCT$.
	\end{itemize}
	\textbf{Organization.} The rest of the paper is organized as follows. In Section 2, we present the model and the problem definition. Section 3 presents an analysis of generating $\IBS$ to estimate the benefit function.
	Our $\IVM$ algorithm along with its theoretical analysis are introduced in Section 4. Experimental results are shown in Section 5. Finally Section 6 concludes the paper.
	\section{Model and Problem Definitions}
	In this section, we present the well-known Independent Cascade ($\IC$) model and the $\CTVM$ problem. The frequently used notations are summarized in Table \ref{tab:notation}. 
	\begin{table}[h]
		\caption{Table of symbols}
		\label{tab_sym}   
		\begin{tabular}{lp{5cm}|lp{4cm}}
			\hline\noalign{\smallskip}
			\textbf{Symbol} & \textbf{Notation} & \textbf{Symbol} & \textbf{Notation}
			\\
			\noalign{\hrule height 1.0pt} \noalign{\smallskip}
			$n, m$ & \# nodes and \# of edges  in $G$.
			&
			$S$ & seed set
			\\
			$\eB(S)$ & The benefit of $S$ & 
			$\hB(S)$ & An estimation of $\eB(S)$
			\\
			$k_{max}$ & $k_{max}=\max\{k| c(S)\leq B, |S|=k\}$ &	$\Cov(R_j, S)$ & $\min\{1, |R_j \cap S|\}$ 
			\\
			$\mu_{min} $ &  $\sum_{v \in S}(1-\gamma(v))\frac{b(u)}{\Gamma}$ 
			&
			$\mu_{max}$ & $\frac{\Phi}{\Gamma}+ \sum_{v \in S}(1-\gamma(v))\frac{b(u)}{\Gamma}$
			\\
			$\rho$ & $\mu_{max}-\mu_{min}$ & 
			$p$ & $\min\{\mu_{max}-\mu_{min}, \mu_{max}+ \mu_{min}- 2\sqrt{\mu_{min} \mu_{max}}\}$
			\\
			$S^*$ & An optimal solution & 	$\OPT$ &  $\eB(S^*)$.
			\\
			$\Gamma$ & $\Gamma=\sum_{u \in V}b(u)$ & 
			$\Phi$ & $\Phi=\sum_{u \in V}\gamma(u) b(u)$
			\\
			$\alpha(\delta)$&$\left( 1- \frac{1}{\sqrt{e}}\right)  (\ln \frac{2}{\delta} )^{1/2}$& $\beta(\delta)$&$ \left( 1-\frac{1}{\sqrt{e}}\right)   \ln \left(\frac{2k_{max}}{\delta}  \binom{n}{k_{0}} \right)^{1/2}$
			\\
			\noalign{\smallskip}\hline
		\end{tabular}
		\label{tab:notation}
	\end{table}
	\\
	In this model, a social network can be abstracted as a
	directed graph  $G=(V, E)$ with a node set $V$ and a directed edge set $E$,  $|V|=n$ and $|E|=m$.  Let $N_{in}(v)$ and $N_{out}(v)$ be the set of in-neighbors and out-neighbor of $v$,  respectively. Each edge $e=(u, v) \in E$ has a probability $p(u, v) \in (0, 1)$ that represents the information transmission from $u$ to $v$. 
	The diffusion process from a seed set $S$ to the rest of the network happens round by round as follows. At step 0, all nodes in $S$ are activated while the rest of the nodes are inactive. At step $t \geq 1$, an active node $u$ in the previous step $t-1$ has a single chance to activate each currently inactive out neighbour node $v$ with the successful probability $p(u, v)$. Once a node becomes activated, it remains
	in that state in all subsequent steps. The influence propagation stops when no more node can be activated. 
	
	Kempe {\em et al.} \cite{kem03} showed $\IC$ model is equivalent to the reachability in a random graph $g$, called \textit{live-edge} or \textit{sample graph}. We generate a sample graph $g$ with the set of nodes be $V_g$ and the set of edges be $E_g$ by: (1) setting $V_g \leftarrow V $, and (2) selecting $e=(u, v) \in E$ into $E_g$ with probability $p(e)=p(u, v)$. 
	The probability to generate $g$ from $G$ is:
	$
	\Pr[g\sim G]= \prod_{e \in E_g} p(e) \cdot \prod_{e \in E \setminus E_g} (1-p(e))
	$
	and the influence spread of $S$ is calculated by:
	\begin{align}
		\eI(S)=\sum_{g \sim G}{\Pr[g\sim G]|R(g, S)|}
		\label{inf_cal}
	\end{align}
	where $R(g, S)$ denotes the set of reachable nodes from $S$ in $g$. 
	In $\CTVM$, each node $u \in V$ has a cost $c(u) > 0$ if it is selected into the seed set $S$ and a benefit $b(u) \geq 0$ if it is activated. The total benefit over all influenced nodes (benefit function) of seed set $S$ is defined as follows:
	\begin{align}
		\eB(S)=\sum_{g \sim G}\Pr[g\sim G]\sum_{u \in  R(g, S)}b(u)
		\label{eq:ben_fun}
	\end{align}
	$\CTVM$ problem is formally defined  as follows.
	\begin{definition}[$\CTVM$]
		Given a social network $G=(V, E, w)$ with a node set $V$ and a directed edge set $E$ under a $\IC$ model.
		Each node $u \in V$ has a selecting cost $c(u) \geq 0$ and a benefit $b(u)$ if $u$ is active. Given a budget $B>0$, find a seed set $S \subset V$ with the total cost $c(S)\leq B$ to maximize $\eB(S)$.
	\end{definition}
	\section{Importance Benefit Sampling}
	In this section, we first recap the Benefit Sampling Algorithm (BSA) to estimate the benefit function \cite{ctvm_infocom,bct_ton}. We then introduce our novel Importance Benefit Sample ($\IBS$) concept along with the algorithm to generate these samples.
	
	Benefit Sampling Algorithm (BSA) \cite{ctvm_infocom} generates Benefit Sample according to the following steps: (1) picking a node $u$ as a source node with probability $\frac{b(u)}{\Gamma}$, (2) generating a sample graph $g$ from $G$ and 3) returning $R_j$ as the set of nodes that can reach $v$ in $g$. 
	Denote $\cR$ as a collection of benefit samples generated by BSA and define a random variable $Y_j=\min\{|R_j \cap A|, 1\}$. Nguyen {\em et al.} \cite{ctvm_infocom} prove the following Lemma to estimate the benefit function:
	\begin{lemma}
		For any set of nodes $S \subset V$, we have: $\eB(S)= \Gamma \cdot \eE[Y_j]$
		\label{lem_est}
	\end{lemma}
	Let $\Omega$ be a set of all benefit samples and $R_j(u)$ be a benefit sample with source node $u$, the probability of generating $R_j(u)$ is 
	\begin{align}
		\Pr[R_j(u) \sim \Omega]=\frac{b(u)}{\Gamma} \sum_{g \in G: u \in R(g, S)}\Pr[g \sim G]
	\end{align}
	We now describe the $\IBS$ and algorithm that generates $\IBS$.
	The main idea of this method is based on the observation that the  sets containing only one node contributes insignificantly in calculating the benefit function. 
	For a source node $u$, assume $\Omega_u$ is set of all benefit samples that has source node $u$. We divide $\Omega_u$ into two components: 
	$\Omega_u^0$ - singular benefit samples which contain only node $u$ ,and $\Omega_u^n$ - importance benefit samples which contain at least two nodes.
	Let $N_{in}(u)=\{v_1, v_2, \ldots, v_l\}$, denote $E_0$ as the event that none of nodes in $N_{in}(u)$ is selected, we have
	$
	\Pr[E_0]=\prod_{v \in N_{in}(u)}(1-p(v, u))
	$.
	The probability of generating an $\IBS$ with source node $u$ is equal to 
	$
	\gamma(u)=1-\Pr[E_0]
	$.
	Denote $E_i$ as the event that $v_i$ is the first selected node, we have:
	\begin{align}
		\Pr[E_i]=p(v_i, u) \cdot \prod_{j=1}^{i-1} (1-p(v_j, u))
		\label{eq:pro_nei}
	\end{align}
	Events $E_0, E_1, \ldots, E_l$ are disjoint sets and $\sum_{i=0}^l\Pr[E_i]=1$. The probability of generating an $\IBS$ that has source node $u$ with $v_i$ is the first selected node is
	\begin{align}
		\Pr[E_i|R_j(u) \in \Omega_u^n]=\Pr[E_i]/\gamma(u)
	\end{align}
	Denote $ \Omega^n$ as the probability spaces of all $\IBS$s, we have:
	\begin{align}
		\Pr[R_j(u) \sim \Omega^n] =\frac{1}{\gamma(u)} \Pr[R_j(u) \sim \Omega]
	\end{align}
	The probability that  $u$ is a source node of an $\IBS$ $R_j$ in $\Omega$ is $\frac{b(u)}{\Gamma} \gamma(u)$. By normalizing factor to fulfill a probability distribution of a sample space, the probability that  $u$ is a source node of an $\IBS$ $R_j$ in $\Omega^n$ is calculated as follows:
	\begin{align}
		\Pr[src(R_j)=u]= \frac{\gamma(u)b(u)}{\sum_{u \in V} \gamma(u)b(u)}=\frac{\gamma(u) b(u)}{\Phi}
		\label{eq:pro_im}
	\end{align}
	\begin{lemma}
		For any $\IBS$ $R_j$, we have $\Pr[R_j \sim \Omega]=\frac{\Phi}{\Gamma}\cdot  \Pr[R_j \sim \Omega^n]$
		\label{lem:rate_sam}
	\end{lemma}
	\begin{proof}
		We have:
		\begin{align}
			\Pr[R_j \sim \Omega]& = \sum_{u \in V} \Pr[\mbox{$u$ is source of $R_j$ in $\Omega$}] \Pr[R_j(u) \sim \Omega]
			\\
			& =  \sum_{u \in V} \frac{b(u)}{\Gamma} \cdot \gamma(u)  \Pr[R_j(u) \sim \Omega^n]
			\\
			&=\frac{\Phi}{\Gamma} \cdot \sum_{u \in V} \frac{\gamma(u)  b(u)}{\Phi}  \Pr[R_j(u) \sim \Omega^n]
			\\
			&= \frac{\Phi}{\Gamma} \cdot \sum_{u \in V} \Pr[\mbox{$u$ is source of $R_j$ in $\Omega^n$}] \Pr[R_j(u) \sim \Omega^n] 
			\\
			&= \frac{\Phi}{\Gamma}\cdot  \Pr[R_j \sim \Omega^n]
		\end{align}
	\end{proof}
	Based on the above above analysis, we propose $\IBA$, an algorithm to generate an $\IBS$, which is depicted in Algorithm \ref{alg:iba}. The algorithm first selects a source node $u$ with a probability according to eq. \eqref{eq:pro_im} (line 1). It next picks the first incoming node to $u$ (line 2).  The rest algorithm is similar to the Importance Influence Sampling Algorithm \cite{importance_sketch}.
	\begin{algorithm}[h]
		\SetNlSty{text}{}{.}
		\KwIn{Graph $G=(V,E)$ under $\IC$ model} 
		\KwOut{A Benefit Important Samples $R_j$}
		Pick a source node $u$ with probability in eq. \eqref{eq:pro_im}
		\\
		Select an in-neighbour  $v_i \in N_{in}(u)$ of $u$ with probability $\Pr[E_i]/\gamma(u)$
		\\
		Initialize a queue $Q=\{v_i\}$ and $R_j={i, v_i}$
		\\
		\For{$t =i+1$ to $l$}
		{
			With probability $p(v_t, u)$: $Q.push(v_t)$ and $R_j \leftarrow R_j\cup \{v_t\}$
		}
		\While{$Q$ is not empty}
		{
			$v \leftarrow Q.pop()$
			\\
			\ForEach{$u \in N_{in}(v) \setminus (R_j \cup Q)$}
			{
				$Q.push(u)$, $R_j \leftarrow R_j \cup \{u\}$
			}
		}
		\Return $R_j$
		\caption{$\IBA$  for $\IC$ model}
		\label{alg:iba}
	\end{algorithm}
	For any $\IBS$ $R_j$ is generated by $\IBA$, we define random variables $ X_j(S) =\min\{1, |S\cap R_j|\}$, and
	\begin{align}
		Z_j(S) =\frac{\Phi}{\Gamma}  \cdot X_j(S)+ \sum_{v \in S}(1-\gamma(v))\frac{b(u)}{\Gamma}
	\end{align}
	We have $Z_j(S) \in [\mu_{min}, \mu_{max}]$, with $\mu_{min}=\sum_{v \in S}(1-\gamma(v))\frac{b(u)}{\Gamma}, \mu_{max}=\frac{\Phi}{\Gamma}+ \sum_{v \in S}(1-\gamma(v))\frac{b(u)}{\Gamma}$.
	The following Lemma shows that we can evaluate benefit function by the expectation of $Z_j(S)$. 
	\begin{lemma}
		For any set of nodes $A \subseteq V$, we have: 
		\begin{align}
			\eB(S)= \Phi \cdot \eE[X_j(S)] + \sum_{v \in S}(1-\gamma(v)) b(u)= \Gamma \cdot \eE[Z_j(S)]
		\end{align}
		\label{lem: ibs_est}
	\end{lemma}
	\begin{proof}
		Let $\Cov(S, R_j)=\min\{1, |R_j \cap A|\}$ and $\Omega^0=\Omega \setminus \Omega^n$. From Lemma \ref{lem_est}, we have $\eB(S)  = \Gamma \cdot \sum_{R_j \in \Omega} \Pr[R_j \sim \Omega] \Cov(S, R_j)$
		\begin{align}
			& = \Gamma \cdot (  \sum_{R_j \in \Omega^0}  \Pr[R_j \sim \Omega]  \Cov(S, R_j) +  \sum_{R_j \in \Omega^n}  \Pr[R_j \sim \Omega] \Cov(S, R_j) )
			\label{eq:lem_pro1}
		\end{align}
		Since each  $R_j \in \Omega \setminus \Omega^n$ contains only source node, $X_j(S)=1$ if $R_j \in S$. In this case, we have $\Pr[R_j \sim \Omega]=\frac{b(u)}{\Gamma}(1-\gamma(u))$, with $u=src(R_j)$. Put it into \eqref{eq:lem_pro1}, we have:
		\begin{align}
			\eB(S) &  = \Gamma  \sum_{u\in S} \frac{b(u)}{\Gamma}(1-\gamma(u))  + \Gamma  \sum_{R_j \in \Omega^n}  \Pr[R_j \sim \Omega] \Cov(S, R_j)
			\\
			&=   \sum_{u\in S} \frac{b(u)}{\Gamma}(1-\gamma(u)) + \Gamma \sum_{R_j \in \Omega^n} \frac{\Phi}{\Gamma} \Pr[R_j \sim \Omega^n] \Cov(S, R_j)
			\\
			& = \sum_{u\in S} \frac{b(u)}{\Gamma}(1-\gamma(u)) + \Phi \eE[Z_j(S)]
		\end{align}
		This completes the proof.
	\end{proof}
	Basically, Lemma \ref{lem: ibs_est}  generalizes the result of Lemma 3 in \cite{importance_sketch} in which important reverse reachable sets  (sketches) can be used to estimate the influence spread.
	Therefore, an estimation $\eB(S)$ over a collection of $\IBS$  $\cR$ is:
	\begin{align}
		\hB(S)=  \frac{\Phi}{|\cR|} \sum_{R_j \in \cR} \Cov(S, R_j)+ \sum_{v \in S}(1-\gamma(v))b(u)= \frac{\Gamma}{|\cR|}\sum_{i=1}^{|\cR|} Z_j(S)
		\label{est2}
	\end{align}
	\section{Importance Sample-based Viral Marketing Algorithm}
	We present  Importance Sample-based Viral Marketing ($\IVM$), an $(1-1/\sqrt{e}-\epsilon)$-approximation algorithm for $\CTVM$. $\IVM$ includes two components: generating $\IBS$ to estimate the benefit function and new strategy to find candidate solution and checks its approximation guarantee condition by developing two \textit{lower} and \textit{upper bound} functions. 
	\\
	\textit{\textbf{Algorithm description.}} Our $\IVM$ algorithm is depicted in Algorithm \ref{alg:IVM}. It first calculates the maximum number of $\IBS$s $N_{max}= N(\epsilon, \frac{\delta}{3}) \cdot \frac{\mathsf{OPT} }{l\OPT}$ (line 1), where $l\OPT$ is  defined as follows
	\begin{align}
		l\OPT=\max_{S \subseteq V, c(S) \leq B}{\sum_{s\in S} b(s)}
	\end{align}
	$l\OPT$ is a lower-bound of $\OPT$,	 which is calculated by Algorithm \ref{alg:lOPT}. 
	\begin{algorithm}[h]
		\SetNlSty{text}{}{.}
		\KwIn{Graph $G=(V,E)$, budget $B>0$} 
		\KwOut{ $l\OPT$}
		$\l\OPT \leftarrow 0$, $S \leftarrow \emptyset, i \leftarrow 1$
		\\
		Sort $V=\{v_1, \ldots, v_n\}$ in descending order $b(v_i)$
		\\
		\While{True}
		{
			\eIf{$c(S)+c(v_i) \leq B$}
			{
				$S \leftarrow S \cup \{v_i\}$;
				$l \OPT \leftarrow l \OPT +  b(v_i)$,
			}{
				break;
			}
		}
		\Return $l\OPT$;
		\caption{Calculate $l\OPT$}
		\label{alg:lOPT}
	\end{algorithm}
	Using $l\OPT$ significantly reduces the number of  required samples while still ensuring  total samples. 
	
	$\IVM$ then generates a set of $\IBS$s $\cR_1$ contains $N_1$ samples (line 2). The main phrase consists of at most $t_{max}=\Big\lceil \log_2 \frac{N_{max}}{N_1} \Big\rceil$ iterations (line 4-14). In each iterator $t$, the algorithm maintains a set $R_t$ consists $N_1 \cdot2^{t-1}$ and finds a candidate solution $S_t$ by using Improve Greedy Algorithm ($\IGA$) for Budgeted Maximum Coverage (BMC) problem \cite{bmc}. The details of $\IGA$ is presented in Algorithm  \ref{greedy}. It	finds solution for instance $(\cR_t, b(u), B)$ in which $\cR_t$ is a set of samples, $V$ is the universal set and $B$ is the budget. This algorithm returns $(1-1/\sqrt{e})$-approximation solution \cite{bmc}. The main algorithm then calculates: $f_l(R_t, \delta_1)$ - a lower bound of $\eB(S_t)$ and  $f_u(\cR_t, \delta_1)$ - a upper bound of optimal value $\OPT$ (line 7). We show that $\Pr[f_l(R_t, \delta_1)\geq \eB(S_t)] \geq 1-\delta_1$ (Lemma \ref{lem:lower_bound}) and $\Pr[f_u(R_t, \delta_1)\geq \OPT] \geq 1-\delta_1$ (Lemma \ref{lem:upper_bound}). The algorithm checks approximation guarantee condition: $\frac{f_l(\cR_t, \delta_1)}{f_u(\cR_t, \delta_1)} \geq 1-1/\sqrt{e}-\epsilon $  (line 8). If this condition is true, it returns  $S_t$ as a solution and terminates. If not, it doubles number of samples (line 12) and moves onto the next iterator $t+1$. 
	
	\begin{algorithm}[H]
		\SetNlSty{text}{}{.}
		\KwIn{A set of samples $\cR, k$.} 
		\KwOut{ $S$}
		$S_1 \leftarrow \emptyset$, $U \leftarrow V$
		\\
		\While{$U\neq \emptyset$}
		{
			$u_{max} \leftarrow \arg \max_{u \in U}(\hB(S\cup\{u\})-\hB(S_1))/c(u)$
			\\
			\If{$c(u_{max}) \leq B- c(S_1)$}
			{
				$S_1 \leftarrow S_1 \cup \{u_{max}\}$, 
			}
			$U \leftarrow U \setminus u_{max}$
		}
		$v_{max} \leftarrow \arg \max_{u\in V| c(u) \leq B}\hB(u)$
		\\
		$S \leftarrow \arg \max{S' \in \{S_1, v_{max}\} } \hB(S')$
		\\
		\Return $S$;
		\caption{Improve Greedy Algorithm  ($\IGA$)}
		\label{greedy}
	\end{algorithm} 
	\begin{algorithm}[h]
		\SetNlSty{text}{}{.}
		\KwIn{Graph $G=(V,E)$, budget $B>0$, and $\epsilon, \delta \in (0, 1)$} 
		\KwOut{ seed $S$}
		$N_{max} \leftarrow N(\epsilon, \frac{\delta}{3}) \cdot \frac{\mathsf{OPT} }{l\OPT}$.
		\\
		$N_1 \leftarrow \frac{1}{ \epsilon^2} \ln \frac{1}{\delta}$, $t \leftarrow 1$,  $N_0 \leftarrow 0$
		\\
		$	t_{max}\leftarrow \Big\lceil \log_2 \frac{N_{max}}{N_1} \Big\rceil$, $\delta_1 \leftarrow \frac{\delta}{3 t_{max}}$
		\\
		\Repeat{$|\mathcal{R}_t| \geq N_{max}$}
		{
			
			Generate more $N_t-N_{t-1}$ $\IBS$s and add them into $\mathcal{R}_t$
			\\
			$ <S_t, \hB(S_t)>  \leftarrow \IGA(\mathcal{R}_t, B)$
			\\
			Calculate $f_l(\cR_t, \delta_1)$ by \eqref{lower_bound_eq} and calculate $f_u(\cR_t, \delta_1)$ by  \eqref{upper_bound_eq}.
			\\
			\eIf{$\frac{f_l(\cR_t, \delta_1)}{f_u(\cR_t, \delta_1)}\geq 1-1/\sqrt{e}-\epsilon$}
			{
				\Return $S_t$	
			}
			{ 
				$t \leftarrow t+1$, $R_t \leftarrow R_{t-1}$
				\\
				$N_t \leftarrow 2 N_{t-1} $
				
			}
		}
		\Return $S_t$;
		\caption{$\IVM$ algorithm}
		\label{alg:IVM}
	\end{algorithm}
	
	\textit{\textbf{Theoretical analysis.}} 
	We observe that $Z_j(S) \in [0, 1]$. Let randomly variable $M_i=\sum_{j=1}^i (Z_j(S) - \mu), \forall i \geq 1$, where $\mu=\eE[Z_j]$. For a sequence random variables $M_1, M_2, \ldots$ we have $\eE[M_i| M_1, \ldots, M_{i-1}]= \eE[M_{i-1}]+ \eE[Z_i(S) - \mu]=\eE[M_{i-1}]$. Hence, $M_1, M_2, \ldots$ be a form of martingale \cite{mar_le}.  We have following result from \cite{mar_le}.
	\begin{lemma}
		If $M_1, M_2, \ldots$ be a form of martingale, $|M_1| \leq a$, $|M_j -M_{j-1}|\leq a$ for $j \in [1, i]$, and 
		\begin{align}
			\Var[M_1]+ \sum_{j=2}^i \Var[M_j| M_1, M_2, \ldots, M_{j-1}] = b
		\end{align}
		where $\Var[\cdot]$ denotes the variance of a random variable. Then, for
		any $\lambda$, we have:
		\begin{align}
			\Pr[M_i - \eE[M_i] \geq \lambda] \geq \exp\left(-\frac{\lambda^2}{\frac{2}{3}a \lambda + 2b} \right) 
			\label{mar_inq}
		\end{align}
		\label{mar}
	\end{lemma}
	Apply Martingale theory \cite{mar_le}, we have the following Lemma:
	\begin{lemma}
		For any $T>0, \lambda>0$, $\mu$ is the mean of $Z_j(S)$, and an estimation of $\mu$ is $\hat{\mu}=\frac{\sum_{i=1}^T Z_i(S)}{T}$, we have:
		\begin{align}
			\Pr\Big[\sum_{j=1}^T Z_j(S) - T \cdot \mu \geq \lambda \Big] & \leq  \exp\left(-\frac{\lambda^2}{\frac{2}{3}(\mu_{max}-\mu_{min})\lambda + 2 p \mu T}\right) 
			\label{mar_lemma_1}
			\\
			\Pr\Big[\sum_{j=1}^T Z_j(S) - T \cdot \mu \geq - \lambda \Big] & \leq  \exp\left(-\frac{\lambda^2}{ 2p\mu T}\right) 
			\label{mar_lemma_2}
		\end{align}
		\label{lem:mar_apply}
	\end{lemma}
	\begin{proof}
		Since $Z_j(S) \in [ \mu_{min}, \mu_{max}]$, we have
		\begin{align*}
			\Var[Z_j(S)] &
			= \eE[\eE[Z_j(S)]-Z_j(S)]^2
			\\
			& \leq ( \mu- \mu_{min}) (\mu_{max}- \mu)= \mu \left( \mu_{max} + \mu_{min} - \left( \mu+ \frac{\mu_{min} \mu_{max}}{\mu}\right)  \right) 
		\end{align*}
		Apply Cauchy's inequality, we have
		$ \mu+ \frac{\mu_{min} \mu_{max}}{\mu} \geq 2 \sqrt{\mu_{min} \mu_{max}} $. Therefore,
		\begin{align}
			\Var[Z_j(S)] 
			& \leq  \mu \left( \mu_{max} + \mu_{min} - 2\sqrt{\mu_{min} \mu_{max}}  \right) 
			\label{cb1}
		\end{align}
		On the other hand
		\begin{align}
			\Var[Z_j(S)] \leq ( \mu- \mu_{min}) (\mu_{max}- \mu) \leq \mu(\mu_{max}-\mu_{min})
			\label{cb2}
		\end{align}
		Combine \eqref{cb1} and \eqref{cb2}, we have $\Var[Z_j(S)] \leq p$.
		Since $|M_1|=|Z_1-\mu| \leq \mu_{max}- \mu_{min}$,  $|M_i-M_{i-1}|=|Z_i-Z_{i-1}| \leq \mu_{max} - \mu_{min}$, and 
		\begin{align}
			\Var[M_1]+ \sum_{j=2}^i \Var[M_j| M_1, M_2, \ldots, M_{j-1}] = \sum_{j=1}^T \Var[T_j(S)]\leq p \mu T 
		\end{align}
		Apply Lemma \ref{mar}, we chose $a=\mu_{max}-\mu_{min}$ and $b=T p \mu$, and put back into \eqref{mar_inq} we obtain \eqref{mar_lemma_1}.
		Similarly, $-M_1,\ldots, -M_i, \ldots$ also form a Martingale and applying Lemma \eqref{mar_inq}, we obtain \eqref{mar_lemma_2}.
	\end{proof}
	Assume that $S$ is a feasible solution of $\CTVM$. We do not known the size of $S$ but we can observe that the number of possible solutions is less than $\sum_{k=1}^{k_{max}}\binom{k}{n} \leq k_{max} \binom{k_0}{n}$, where $k_0=\arg \max_{k=1 \ldots k_{max}}\binom{k}{n}$.  Similar result to Theorem 1 in \cite{tang15}. With some modifications, we have following result.
	\begin{theorem}
		For $ \epsilon >0$, $\delta \in (0, 1)$. If the number of samples $|\cR| \geq N(\epsilon,\delta)=\frac{2  \rho  \Gamma \epsilon^{-2}}{\OPT } \left( \alpha(\delta)+\beta(\delta) \right)^{2} 
		$, $\IGA$ returns a ($1-1/\sqrt{e}-\epsilon$)-approximation solution with probability at least $1-\delta$.
		\label{numofsam}
	\end{theorem}
	We can apply Theorem 
	\ref{numofsam} to obtain the following Corollary:
	\begin{coro} At iterator $t$,
		$\Pr[(N_t\geq N_{max})\wedge(\eB(S_t) < (1-\frac{1}{\sqrt{e}}-\epsilon)\OPT)] \geq 1-\delta_1$
		\label{coro:nmax}
	\end{coro}
	\begin{proof}
		$N_{max}=N(\epsilon, \delta/3)\cdot \frac{l \OPT}{\OPT}\geq N(\epsilon, \delta/3)$. Since $S_t$ is returned by $\IGA$ algorithm, apply Theorem 
		\ref{numofsam}, we obtain the proof.
	\end{proof}
By using Lemma \ref{lem:mar_apply}, we give two lower-bound and upper-bound functions in Lemma \ref{lem:lower_bound} and Lemma \ref{lem:upper_bound}. They help the main algorithm check the approximate condition of the candidate based on statistical evidence.
	\begin{lemma}[Lower-bound function] For any $\delta \in (0, 1)$, a set of $\IBS$s $\cR$ and $\hB(S)$ is an estimation of $\eB(S)$ over $\cR$ by \eqref{est2}.	 Let $c=\ln (\frac{1}{\delta})$ and
		\begin{align}
			f_l( \cR, \delta)=  \min \left\{ \hat{\eB}(S)-\frac{\rho c \Gamma}{3T},\hB(S)- \frac{\Gamma}{T}\left( \frac{\rho c}{3}- cp + \sqrt{\left( \frac{\rho c}{3}- cp\right)^2+ 2 Tp c\frac{\hB(S)}{\Gamma}}\right) \right\}
			\label{lower_bound_eq}
		\end{align}
		we have $\Pr[\eB(S) \geq \hB(S)] \geq 1-\delta$
		\label{lem:lower_bound}
	\end{lemma}
	\begin{proof}
		From  inequality \eqref{mar_lemma_1} in Lemma \ref{lem:mar_apply},  let $\lambda=\frac{\rho c}{3}+ \sqrt{\frac{\rho^2c^2}{9}+ 2 c \mu T p}$, we have:
		\begin{align}
			\Pr\Big[\sum_{j=1}^T Z_j(S) - T \cdot \mu \geq \lambda \Big] & \leq  \delta
		\end{align}
		Therefore, the following event happens with probability at least $1-\delta$
		\begin{align}
			\sum_{j=1}^T Z_j(S) - T \cdot \mu \leq \lambda
			\Leftrightarrow  T \hat{\mu} - T \mu -\frac{\rho c}{3} \leq  \sqrt{\frac{\rho^2c^2}{9}+ 2 c \mu T p} \label{bpt}
		\end{align}
		We consider two following cases: (1) If $T \hat{\mu} - T \mu -\frac{\rho c}{3} \leq 0$, then $\mu \geq \hat{\mu} -\frac{\rho c}{3T}$, (2) if $T \hat{\mu} - T \mu -\frac{\rho c}{3} > 0$, \eqref{bpt} becomes:
		\begin{align}
			& \left( T \hat{\mu} - T \mu -\frac{\rho c}{3}\right)^2 \leq  \frac{\rho ^2c^2}{9}+ 2 c \mu T p
			\\
			\Leftrightarrow \   & (\hat{\mu}-\mu)^2 T - 2\left(\frac{\rho c}{3}-cp \right)(\hat{\mu}-\mu) - 2 c \hat{\mu} p \leq 0
		\end{align}
		Solve the above  inequality for $\mu$, we obtain:
		\begin{align}
			\mu \geq \hat{\mu} - \frac{1}{T}\left( \frac{\rho c}{3}- cp + \sqrt{\left( \frac{\rho c}{3}- cp\right)^2+ 2 Tp c \hat{\mu}}\right)
		\end{align}
		Note that $\mu=\frac{\eB(S)}{\Gamma}, \hat{\mu}=\frac{\hB(S)}{\Gamma}$, we have:
		\begin{align}
			\eB(S) \geq \hB(S) - \frac{\Gamma}{T}\left( \frac{\rho c}{3}- cp + \sqrt{\left( \frac{\rho c}{3}- cp\right)^2+ 2 Tp c \frac{\hB(S)}{\Gamma}}\right)
		\end{align}
	\end{proof}
	\begin{lemma}[Upper-bound function] For any $\delta \in (0, 1)$, a set of $\IBS$s $\cR$, $S_G$ is a solution return by $\IGA$ for input data $(\cR, B)$, and $\hB(S_G)$ is an estimation of $\eB(S)$ over $\cR$ by \eqref{est2}. Let
		\begin{align}
			f_u(\cR, \delta)=  \frac{\hB(S_G)}{1-1/\sqrt{e}} +\frac{\Gamma}{T}\left(-cp+\sqrt{c^2 p^2 + 2Tcp \frac{\hB(S_G)}{(1-1/\sqrt{e})\Gamma} } \right)
			\label{upper_bound_eq}
		\end{align}
		we have $\Pr[\OPT \leq f_u(\cR, \delta)] \geq 1-\delta$
		\label{lem:upper_bound}
	\end{lemma}
	\begin{proof}
		From  inequality \eqref{mar_lemma_1} in Lemma \ref{lem:mar_apply},  let $\lambda=\frac{\rho c}{3}+ \sqrt{\frac{\rho^2c^2}{9}+ 2 c \mu T p}$, we have:
		\begin{align}
			\Pr\Big[\sum_{j=1}^T Z_j(S) - T \cdot \mu \geq \lambda \Big] & \leq  \delta
		\end{align}
		Therefore, the following event happen with prop ability at least $1-\delta$
		\begin{align}
			\sum_{j=1}^T Z_j(S) - T \cdot \mu \leq \lambda
			\Leftrightarrow  T \hat{\mu} - T \mu -\frac{\rho c}{3} \leq  \sqrt{\frac{\rho^2c^2}{9}+ 2 c \mu T p} \label{bpt}
		\end{align}
		We consider two following cases: (1) If $T \hat{\mu} - T \mu -\frac{\rho c}{3} \leq 0$, then $\mu \geq \hat{\mu} -\frac{\rho c}{3T}$, (2) if $T \hat{\mu} - T \mu -\frac{\rho c}{3} > 0$, \eqref{bpt} becomes:
		\begin{align}
			& \left( T \hat{\mu} - T \mu -\frac{\rho c}{3}\right)^2 \leq  \frac{\rho ^2c^2}{9}+ 2 c \mu T p
			\\
			\Leftrightarrow \   & (\hat{\mu}-\mu)^2 T - 2\left(\frac{\rho c}{3}-cp \right)(\hat{\mu}-\mu) - 2 c \hat{\mu} p \leq 0
		\end{align}
		Solve the above  inequality for $\mu$, we obtain:
		\begin{align}
			\mu \geq \hat{\mu} - \frac{1}{T}\left( \frac{\rho c}{3}- cp + \sqrt{\left( \frac{\rho c}{3}- cp\right)^2+ 2 Tp c \hat{\mu}}\right)
		\end{align}
		Note that $\mu=\frac{\eB(S)}{\Gamma}, \hat{\mu}=\frac{\hB(S)}{\Gamma}$, we have:
		\begin{align}
			\eB(S) \geq \hB(S) - \frac{\Gamma}{T}\left( \frac{\rho c}{3}- cp + \sqrt{\left( \frac{\rho c}{3}- cp\right)^2+ 2 Tp c \frac{\hB(S)}{\Gamma}}\right)
		\end{align}
	\end{proof}
	\begin{theorem}[Approximation ratio] $\IVM$ returns $(1-1/\sqrt{e}-\epsilon)$-approximation solution algorithm with probability at least $1-\delta$.
	\end{theorem}
	\begin{proof} We consider following events
		\begin{itemize}
			\item[1)] $E_1(t): f_l(\cR_t, \delta_1) > \eB(S_t)$
			\item[2)] $E_2(t): f_u(\cR_t, \delta_1) < \OPT$
			\item[3)] $E_3: (|\cR_t|\geq N_{max}) \wedge (\eB(S_{t_{max}}) <   (1-\frac{1}{\sqrt{e}}-\epsilon)\OPT) $
		\end{itemize}
		According to Lemmas \ref{lem:lower_bound}, \ref{lem:upper_bound}, and Corollary \ref{coro:nmax}, we have: $\Pr[E_1(t)] \leq \delta_1, \Pr[E_2(t)] \leq \delta_1$ and $\Pr[E_3] \leq \delta/t_{max}$.  Apply the union bound the probability that none of  events  $E_1(t), E_e(t), E_3, \forall t=1, \ldots, t_{max}$ at least $1- \left( \delta_1 \cdot t_{max}+\delta_1 \cdot t_{max} + \frac{\delta}{3} \right) =1- \delta
		$. Under this assumption, we will show that $\IVM$  returns a solution satisfying $
		\eB(S_t) \geq (1-1/\sqrt{e}-\epsilon) \OPT
		$. If the algorithm stops with condition $|\mathcal{R}_t| \geq  N_{max}$, the solution $S_t$ satisfies approximation guarantee due to Corollary \ref{coro:nmax}. Otherwise, if $\IVM$ stops at some iterator $t, t=1,2, \ldots, t_{max}$. At this iterator, the condition in line 8 is satisfied, i.e,
		\begin{align}
			\frac{\eB(S_t)}{\OPT} \geq \frac{f_l(\cR_t, \epsilon_1)}{f_u(\cR_t, \epsilon_1)} \geq 1-1/\sqrt{e}-\epsilon
		\end{align}
		This completes the proof
	\end{proof}
	\begin{theorem}[Number of samples]
		Given $0< \epsilon<1, 0<\delta < \frac{1}{2}$, the expected number of required samples generated by Algorithm \ref{alg:IVM} is  $O( \frac{\rho n\log(M/\delta)}{\epsilon^2 \OPT})$, where $M=k_{max}\binom{n}{k_0}$. 
		\label{theo:numsam}
	\end{theorem}
	\begin{proof}
		For each iterator $t$, let $N_t=|\cR_t|=|\cR_t^c|$,
		\\
		$\epsilon_1= \max \left\{\frac{\rho c \Gamma}{3N_t},\frac{\Gamma}{N_t}\left( \frac{\rho c}{3}- cp + \sqrt{\left( \frac{\rho c}{3}- cp\right)^2+ 2 N_tp c\frac{\hB(S_t)}{\Gamma}}\right)\right\}\frac{1}{\OPT}$, $ \epsilon_2=\frac{\epsilon_4}{3}$, 
		$z=\frac{\Gamma}{T}\left(-cp+\sqrt{c^2 p^2 + 2N_tcp \frac{\hB(S_t)}{(1-1/\sqrt{e})\Gamma} } \right)$, $\epsilon_3=z/(z+\frac{\hB(S_t)}{1-1/\sqrt{e}})$, $\epsilon_4=\frac{\epsilon}{2(1-\frac{1}{\sqrt{e}})}$, and 
		\begin{align*}
			N=\max \left\lbrace \frac{6 \rho \Gamma \log( 2 M/\delta)}{\epsilon_2^2 \OPT},\frac{(\frac{2}{3}\rho + 2p) \Gamma \log(2 M/\delta) }{\epsilon_2^2 \OPT}\right\rbrace 
		\end{align*}
		We have $N=O(\frac{\rho n\log(M/\delta)}{\epsilon^2 \OPT})$. Assume that $N_t=C N$, for any $C >1$.
		Apply Lemma \ref{lem:mar_apply} with $\lambda=\frac{\epsilon_2 \eB(S_t) N_t}{\Gamma}$, we have:
		\begin{align}
			&\Pr[\hB(S_t)\leq (1-\epsilon_2)\eB(S_t)] \leq \exp(-\epsilon_2^2\frac{\eB(S_t)}{2p \Gamma} N_t) \nonumber
			\\
			& \leq \exp(\frac{ \rho \eB(S_t) }{p \OPT} C \ln \frac{\delta}{2M})  \leq  \exp(C \ln \frac{\delta}{2M}) \nonumber
			\\
			& \leq (\frac{\delta}{2M})^C
			\label{event:e_2}
		\end{align}
		And
		\begin{align}
			&\Pr[\hB(S_t)\geq  (1+ \epsilon_2)\eB(S_t)] \leq \exp\left(\frac{-\epsilon_2^2\eB(S_t) N_t}{(\frac{2}{3} \rho+ 2p )\Gamma} \right) \nonumber
			\\
			& \leq \exp(C \ln \frac{\delta}{2M}) \leq (\frac{\delta}{2M})^C
			\label{event:e_3}
		\end{align}
		Under the assumption that event \eqref{event:e_3} do not happen, the following event happens with probability at least $1- (\frac{\delta}{2M})^C$:
		\begin{align*}
			& \frac{\Gamma}{N_t \OPT}\left(\frac{\rho c}{3}- cp + \sqrt{\left( \frac{\rho c}{3}- cp\right)^2+ 2 N_tp c \frac{\hB(S_t)}{\Gamma}}\right)
			\\
			& \leq \frac{1}{ N_t \mu^*}\left(\frac{\rho c}{3} + \sqrt{( \frac{4 \rho c}{3})^2+ 2 N_t \rho  c \frac{(1+\epsilon_2)\eB(S_t)}{\Gamma} } \right)
			\\ 
			& \leq \frac{1}{N_t \mu^* } \left(\frac{\rho c}{3} + \frac{4\rho c}{3} + \sqrt{2 N_t \rho  c \mu}\right) 
			\\
			&= \frac{1}{N_t \mu^* } \left(\frac{5 \rho c}{3} + \sqrt{2 N_t \rho  c \mu^*}\right) < \frac{\epsilon_4}{3}
		\end{align*}
		The last inequality due to $N_t \geq \frac{4 \rho \Gamma \log( 2 M/\delta)}{(\frac{\epsilon_4}{3})^2 \OPT}$. Therefore, $0<\epsilon_1\leq \epsilon_2$. 
		When the event in \eqref{event:e_2} do not happen, from definition of $f_l(\cR_t^c, \delta)$, we have:
		\begin{align}
			f_l(\cR_t^c, \delta) & \geq \hB(S_t) -\epsilon_1 \OPT \geq \hB(S_t) -\epsilon_1 \eB(S_t) 
			\\ & \geq (1-\epsilon_1-\epsilon_2) \eB(S_t)
			\label{iq:fl}
		\end{align}
		On the other hand, apply Lemma \ref{lem:mar_apply} with $\lambda=\frac{\epsilon_4 \eB(S_t) N_t}{\Gamma}$,
		\begin{align}
			& \Pr[\hB(S_t)  \geq (1+ \epsilon_4) \eB(S_t)]
			\\
			&
			=\Pr[\hB(S_t) - \epsilon_4 \eB(S_t) \geq  \eB(S_t)] 
			\\
			&
			\leq \exp(\frac{-\epsilon_4^2\eB(S_t) N_t}{(\frac{2}{3} \rho+ 2p )\Gamma} )
			\leq \exp(C \ln \frac{\delta}{2M}) \leq \left( \frac{\delta}{2M}\right) ^C
			\label{event:e_4}
		\end{align}
		When the event in \eqref{event:e_4} do not happen, we have:
		\begin{align}
			\epsilon_3 & \leq  \frac{z}{\frac{\hB(S_t)}{1-1/\sqrt{e}}} \leq   \frac{z}{\hB(S^*)} \leq \frac{z}{(1-\epsilon_4)\OPT} 
			\\
			&\leq \frac{1}{1-\epsilon_4}\frac{\Gamma}{N_t \OPT} \sqrt{c^2 \rho^2 + 2N_tc\rho \frac{\hB(S_t)}{(1-1/\sqrt{e})\Gamma} }
			\\
			&\leq \frac{1}{1-\epsilon_4}\frac{1}{N_t \mu^*} \sqrt{c^2 \rho^2 + 2N_tc\rho \frac{(1+\epsilon_4)\mu}{1-1/\sqrt{e}} }
			\\
			&\leq \frac{1}{1-\epsilon_4}\frac{1}{N_t \mu^*} \sqrt{c^2 \rho^2 +  \frac{4N_tc\rho\mu}{1-1/\sqrt{e}} } 
			\\
			& < \frac{\epsilon}{1-\epsilon_4}  < \frac{\epsilon_4}{3}
		\end{align}
		The last inequality due to $N_t > \frac{3-1/\sqrt{e}}{1-1/\sqrt{e}} \frac{\rho \Gamma \log( 2 M/\delta)}{(\frac{\epsilon_4}{3})^2 \OPT}$. From definition of $f_u(\cR_t^c, \delta)$, we have:
		\begin{align}
			&f_u(\cR_t^c, \delta) = \frac{\hB(S_t)}{1-1/\sqrt{e}} + z
			\\
			& =  \frac{\hB(S_t)}{1-1/\sqrt{e}} +  \frac{\epsilon_3 \hB(S_t)}{(1-1/\sqrt{e})(1-\epsilon_3)}
			\\
			& = \frac{ \hB(S_t)}{(1-1/\sqrt{e})(1-\epsilon_3)} 
			\label{iq:fu}
		\end{align}
		Since there are at most $M$ candidate solutions $S_t$, By the union bound, the probability that none of the events  \eqref{iq:fl} and \eqref{iq:fu} happens is at least 
		\begin{align}
			1 - (M  (\frac{\delta}{2M})^C +  M (\frac{\delta}{2M})^C) \leq 1- \delta^C
		\end{align}
		Combine \eqref{iq:fl} with \eqref{iq:fu}, we have
		\begin{align*}
			&\frac{f_l(\cR_t, \delta)}{f_u(\cR_t^c, \delta)}  \geq \left(  1-\frac{1}{\sqrt{e}}\right)  \frac{(1-\epsilon_1-\epsilon_2)(1-\epsilon_3) \eB(S_t)}{\hB(S_t)}
			\\ & \geq  \left(  1-\frac{1}{\sqrt{e}}\right)  (1-\epsilon_1-\epsilon_2-\epsilon_3) \frac{\eB(S_t)}{\hB(S_t)}
			\\
			& \geq  \left(  1-\frac{1}{\sqrt{e}}\right)  (1-\epsilon_4) \frac{\hB(S_t) - \epsilon_4 \eB(S_t)}{\hB(S_t)}
			\\
			&\geq  \left(  1-\frac{1}{\sqrt{e}}\right)  (1-\epsilon_4) \left(1- \frac{\epsilon_4 \eB(S_t)}{\hB(S_t)}\right) 
			\\
			& \geq  \left(  1-\frac{1}{\sqrt{e}}\right)  (1-\epsilon_4) \left(1- \frac{\epsilon_4 }{1-\epsilon_4}\right) 
			\\
			& =   1-\frac{1}{\sqrt{e}} - \epsilon
		\end{align*}
		Therefore, when $N_t=C N$, the algorithm meets stopping condition (line 8)    with probability at least $1-\delta^C$. Let $i$ is be the fist iteration that the number of required samples reaches $N$. From this iteration onward, the expected  number of required samples further generated is at most:
		\begin{align*}
			\sum_{j \geq i} N_1 2^{i} 2^{j-i} \delta^{C2^{j-i}}&= 2^i N_1  \sum_{j \geq 0}  2^{j} \delta^{C2^{j}}
			\\ 
			& \leq 2^i N_1  \sum_{j \geq 0}  2^{j - 2^j} \ (\mbox{due to $\delta <\frac{1}{2}, C>1$})
			\\
			& \leq 2^i N_1  \sum_{j \geq 0}  2^{-j} \ (\mbox{due to $2^j \geq 2j$})
			\\
			& \leq 4 N  \ (\mbox{due to $2^{i-1} N_1 \leq N$})
		\end{align*}
		Therefore, the expected number of samples is less than $N+4N=5N$.  Hence,  the expected number of required samples is  $O( \frac{\rho n\log(M/\delta)}{\epsilon^2 \OPT})$. 
	\end{proof}
	\section{Experiment}
	In this section, we briefly conduct experiments to compare
	the performance of our algorithm $\IVM$ to other algorithms for $\CTVM$ on for aspects: the solution quality, running time, number of required samples and used memory.
	\subsection{Experimental Settings}
	\textit{Datasets.} We select a diverse set of 4 datasets including Gnutella, Epinion Amazon and DBLP. The description used datasets is provided in Table \ref{tab:dataset}.
	\begin{table}[h]
		\caption{Dataset}
		\begin{center}
			\begin{tabular}{p{2cm}p{1.5cm}p{1.5cm}ccc}
				\hline\noalign{\smallskip}
				\textbf{Dataset} & \textbf{\#Node }&\textbf{ \#Edge} & \textbf{Type} & \textbf{Avg. degree} & \textbf{Source}
				\\
				\hline
				\textbf{Gnutella} & 6.301 & 20.777 & directed & 3.3 & \cite{gnutella}
				\\
				\textbf{Epinion} & 75.879 & 508.837 & directed & 6.7 & \cite{epinions}
				\\
				\textbf{Amazon} & 262.111 & 1.234.877 & directed & 4.7 & \cite{amazon}
				\\
				\textbf{DBLP} &	317.080 & 1.049.866 & undirected & 3.21 & \cite{DBLP}
				\\ 
				\noalign{\smallskip}\hline
			\end{tabular}
		\end{center}
		\label{tab:dataset}
	\end{table}
	\\
	\textit{Algorithms compared.} We compare the $\IVM$ algorithm against the $\BCT$ algorithm \cite{ctvm_infocom}, and two baseline algorithms: Random and Degree.
	\\
	\textit{Parameter setting.} We follow previous works on  $\CTVM$  and $\IM$ \cite{ctvm_infocom,bct_ton,tang15} to set up parameters. The transmission probability $p(u, v)$ is randomly selected in $\{0.001, 0.01, 0.1\}$ according to the Trivalency model.  The cost of a node proportional to the out-degree \cite{ctvm_infocom}: $c(u)=n |N_{out}(u)|/\sum_{v \in V}|N_{out}(v)|$.
	In all the experiments, we choose a random $p = 20\%$ of all the nodes to be the target set and assign benefit 1 and we set $\epsilon= 0.1$ and $ \delta= 1/n$ as
	a default setting. The budget $B$ varies from 1 to 1000. 
	\subsection{Experiment results}
	\begin{figure}[h]
		\begin{center}
			{\includegraphics[width=.4\linewidth]{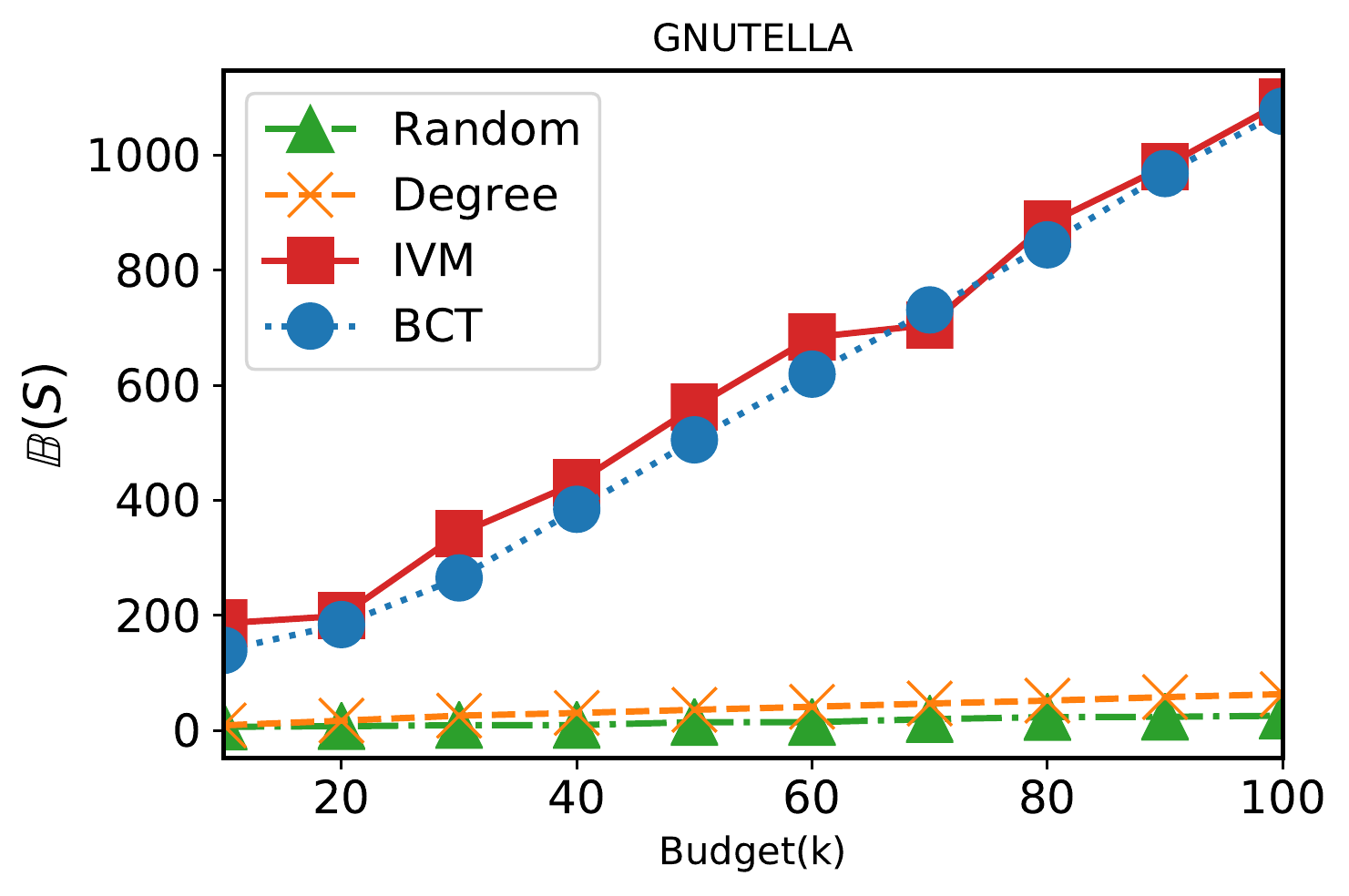}} 
			{\includegraphics[width=.4\linewidth]{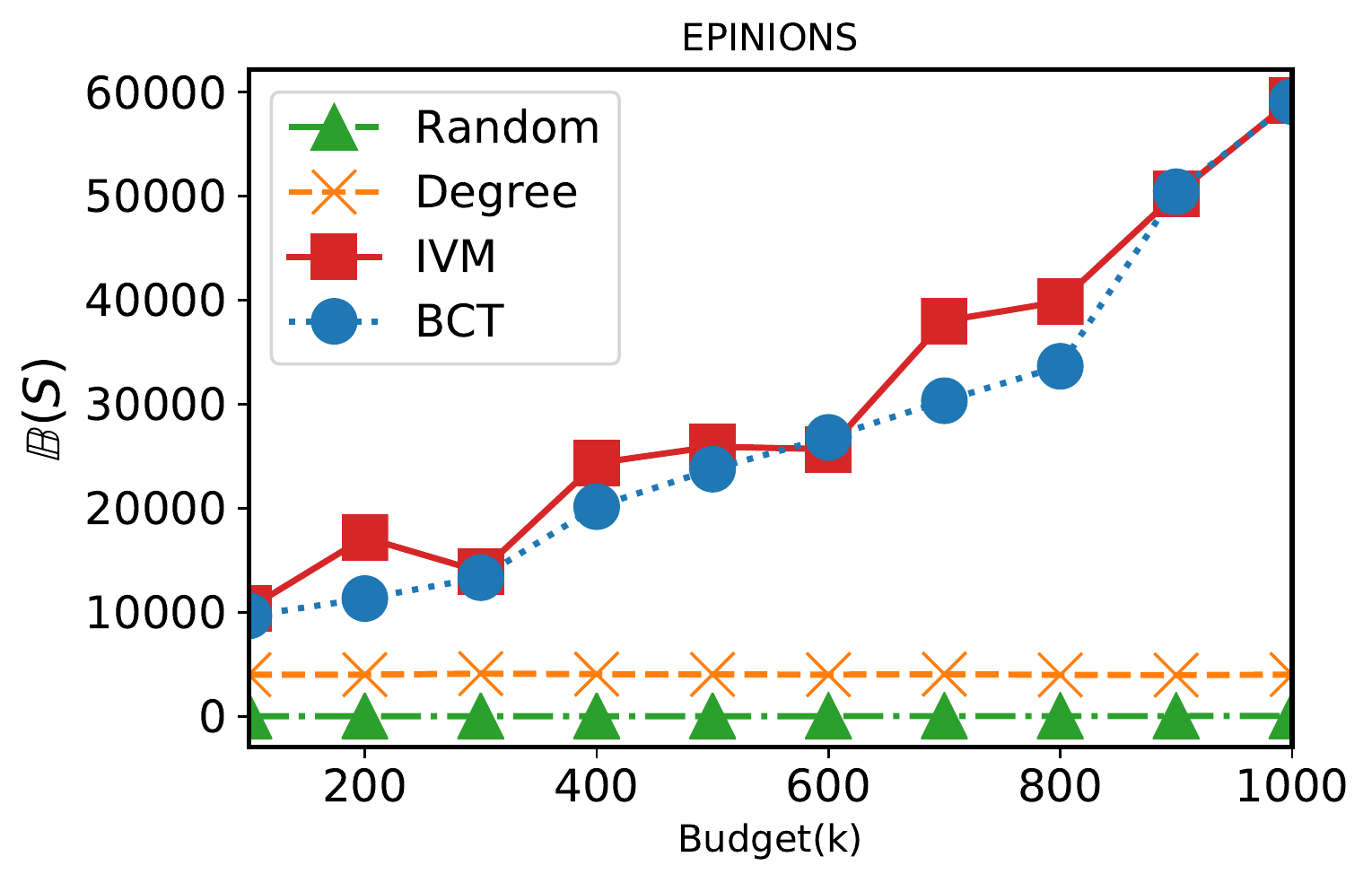}}
			\\
			{\includegraphics[width=.4\linewidth]{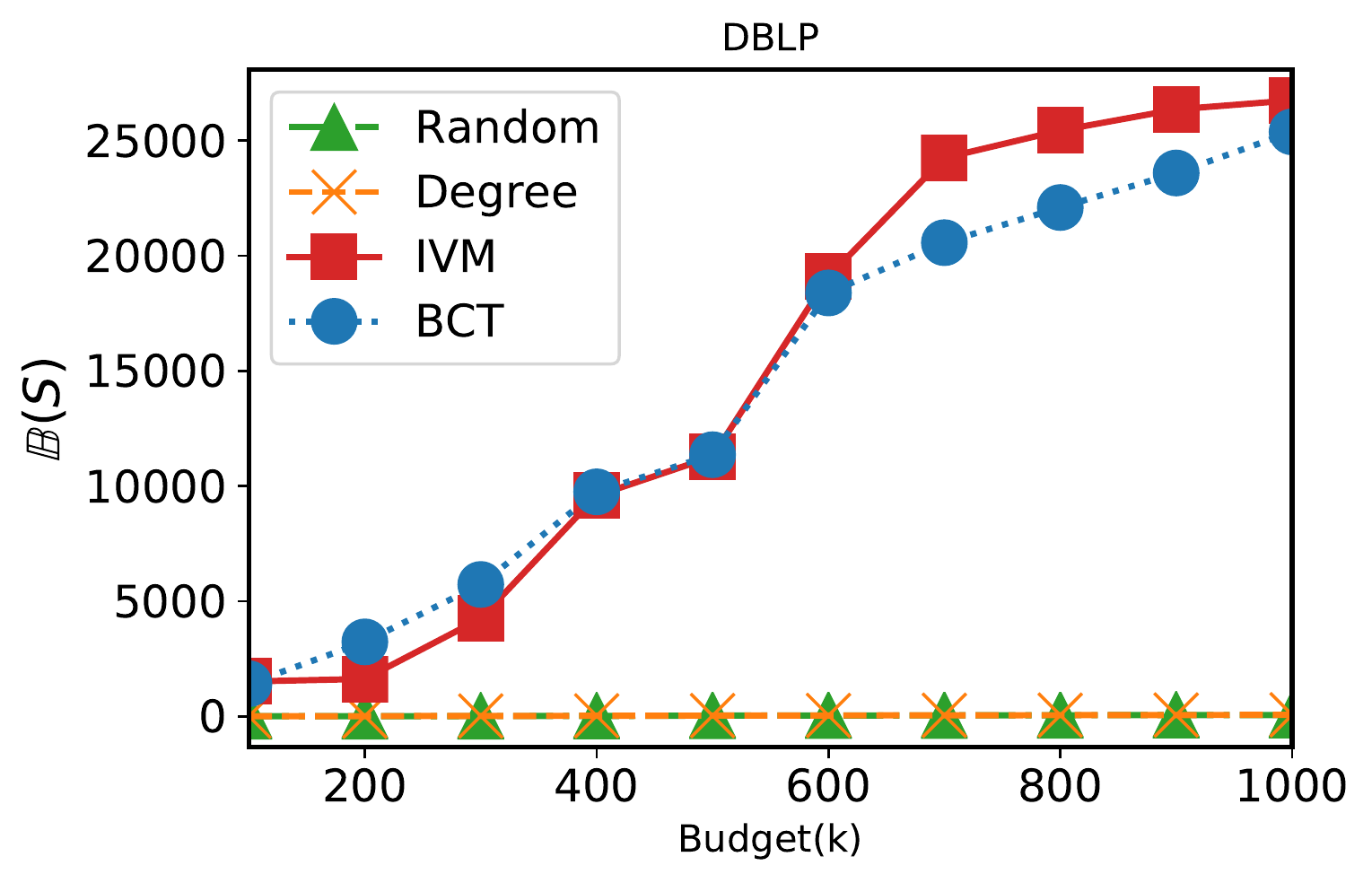}}
			{\includegraphics[width=.4\linewidth]{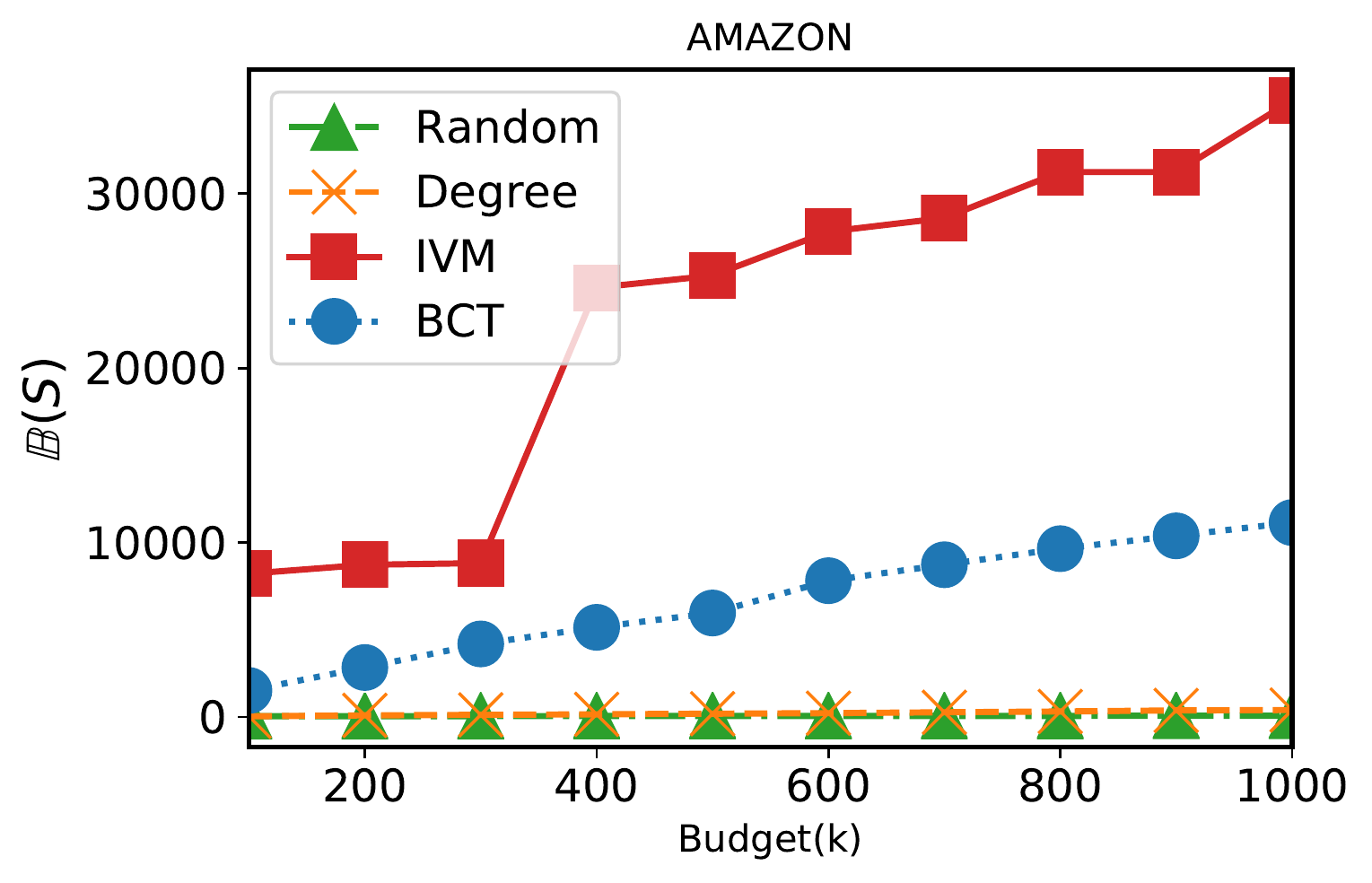}}
		\end{center}
		\caption{The benefit function achieved by algorithms}
		\label{fig:ben-gen-cost}
	\end{figure}
	Fig. \ref{fig:ben-gen-cost} shows the benefit value provided by algorithms. $\IVM$ outperforms other algorithms and gives the best result on Amazon network. It provides up to 5.4 times better than $\BCT$ on Amazon. For Gnutella, Epinions and DBLP networks gives similar result to $\BCT$. This is because these two algorithms give the same approximation ratio for $\CTVM$. 
	\begin{table}
		\caption{Running time between $\IVM$ and $\BCT$ (sec.)}
		\begin{tabular}{c|ccccccccccc}
			\toprule
			\multirow{2}{*}{Network} & & \multicolumn{10}{c}{Budget $B$}\\
			\hhline{~~~~------~~}
			& &100 & 200 & 300 & 400 & 500 & 600 & 700 & 800 & 900 & 1000\\
			\midrule
			\multirow{2}{*}{\textbf{Gnutella}} & \textsf{$\IVM$} & $4.10^{-3}$ & $6.10^{-3}$ & $2.10^{-3}$ & $3.10^{-3}$ & $2.10^{-3}$ & $3.10^{-3}$ & 0.01 & $7.10^{-3}$ & $9.10^{-3}$ & $7.10^{-3}$\\
			& \textsf{$\BCT$} & 0.02 & 0.015 & 0.02 & 0.016& 0.018 & 0.02 & 0.022 & 0.021 & 0.01 & 0.01
			\\
			\hline
			\multirow{2}{*}{\textbf{Epinion}} & \textsf{$\IVM$} & 1.09 & 1.4 & 0.9 & 1 & 1.1 & 1 & 1 & 0.9 & 0.87 & 0.9\\
			& \textsf{$\BCT$} & 7.8 & 0.95 & 6.7 & 3.1 & 3.6 & 3.4 & 3.4 & 3.5 & 1.1 & 1\\
			\hline
			\multirow{2}{*}{\textbf{Amazon}} & \textsf{$\IVM$} & 0.01 & 0.01 & 0.01 & 0.01 & 0.012 & 0.012 & 0.01 & 0.03 & 0.04 & 0.03\\
			& \textsf{$\BCT$} & 1.73 & 0.31 & 0.89 & 0.50 & 0.49 & 0.49 & 0.31 & 0.27 & 0.32 & 0.4\\
			\hline
			\multirow{2}{*}{\textbf{DBLP}} & $\IVM$ & 1.7 & 0.14 & 0.8 & 0.4 & 0.2 & 0.23 & 0.13 & 0.13 & 0.14 & 0.14 \\
			& \textsf{$\BCT$} & 2.6 & 0.4 & 1.7 & 1.9 & 1.1 & 1.2 & 0.7 & 0.6 & 0.5 & 0.4 \\
			\bottomrule
		\end{tabular}
		\label{tab:time}
	\end{table}
	\begin{table}
		\caption{Number of samples and total memory between $\IVM$ and $\BCT$ for  $B=1000$}
		\begin{center}		
			\begin{tabular}{c|cccc|cccc}
				\toprule
				\multirow{2}{*}{Algoirthm} &  \multicolumn{4}{c}{\textbf{Total samples ($\times 10^3$)}} & \multicolumn{4}{c}{\textbf{ Memory usage (M)}}
				\\
				\cline{2-9}
				& \textbf{Gnutella} & \textbf{Epinion} & \textbf{Amazon} & \textbf{DBLP} & \textbf{Gnutella} & \textbf{Epinion} & \textbf{Amazon}  & \textbf{DBLP}
				\\
				\hline
				$\IVM$  & \textbf{0.99} & \textbf{1.12} & \textbf{1.25}& \textbf{1.27} & \textbf{5.9} & \textbf{46} &\textbf{53}& \textbf{66}
				\\
				$\BCT$ & 10 & 10 & 270 & 140 & 22& 67 & 95 & 102
				\\
				\bottomrule
			\end{tabular}
			\label{tab:mem}
		\end{center}
	\end{table}

	The running time of algorithms is shown in Table \ref{tab:time}. The
	running time of our algorithm in all networks are significantly
	lower than  that of $\BCT$. $\IVM$ is up to
	6.4, 7.1, 153  and 4.8 times faster than $\BCT$ on Gnutella, Epinion, Amazon and DBLP networks.

	Table  \ref{tab:mem} displays the memory usage and the number
	of required samples of $\IVM$ and $\BCT$ when the budget $B=1000$.  The number of samples
	generated by $\IVM$ is up to more 112 times smaller than
	that of $\BCT$. However, the memory usage of $\IVM$ is only 1.5 to 4.6 times smaller
	than those of $\BCT$ because of the memory
	for storing the graph is counted into the memory usage of each algorithm. This results  also confirm our theoretical establishment in Section 4 that $\IVM$ requires much less number of samples needed. 
	\section{Conclusion}
	In this paper, we propose $\IVM$, an efficient approximation algorithm for $\CTVM$, which has an approximation ratio of $1-\frac{1}{\sqrt{e}}-\epsilon$ and the expected number of required samples is $O( \frac{\rho n\log(M/\delta)}{\epsilon^2 \OPT})$, which is significantly lower than that of the state-of-the-art $\BCT$. Experiments show that $\IVM$ is up to 153 times faster and  requires up to 112 times fewer total samples than the $\BCT$ algorithm. For the future work, we plan to implement this importance sampling concept on the exact approach $\TIPTOP$ to evaluate potential benefits of the importance sampling in terms of running time and number of required samples.  
	
	%
	\bibliographystyle{splncs04}
	\bibliography{ref}
\end{document}